# A comprehensive protocol for manual segmentation of the human claustrum and its sub-regions using high-resolution MRI

Seung Suk Kang, Joseph Bodenheimer, and Kayley Morris

Department of Biomedical Sciences, University of Missouri-Kansas City, Kansas City, MO, USA.

Tracey Butler

Department of Neuropsychiatry, Weill Cornell Medical College, NY, NY, USA.

Corresponding author: kangseung@umkc.edu

## Abstract

The claustrum (Cl) is a thin grey matter structure located in the center of each brain hemisphere. Cl has been hypothesized as a central hub of the brain for multisensory/sensorimotor integration, consciousness, and attention. Accumulating evidence has suggested that Cl might be important in the development of severe neurological and psychiatric symptoms including epileptic seizures and psychosis. However, the specifics of the roles of Cl in human epilepsy and psychosis are largely unknown, primarily due to methodological limitations related to the thin morphology of Cl that is challenging to delineate accurately using conventional methods. The goal of this work is to develop noninvasive multimodal neuroimaging methods to delineate Cl anatomy by utilizing a large healthy adult high resolution (0.7mm$^3$) T1-weighted MRI collected as part of the Washington University-Minnesota Consortium Human Connectome Project (WU-Minn HCP). We developed a comprehensive manual segmentation protocol to delineate Cl based on a cellular level brain atlas. The protocol involves detailed guidelines to delineate the three subregions of Cl, including the dorsal, ventral, and temporal Cl that can be parcellated based on a geometric method. As demonstrated in a representative result, Cl is large in its anterior-posterior, and the dorsal-ventral extent. Also, the volume is comparable to that of the amygdala. It is required to assess the reliability of the protocol so that it can be used for future anatomical studies of neuropsychiatric disorders, including epilepsy and schizophrenia.

## Introduction

The claustrum (Cl) is a thin deep-brain nucleus in the basolateral telencephalon of the mammalian brain. It is located at the center of each hemisphere and has been characterized as the brain's most highly connected hub (Torgerson, Irimia, Goh, & Van Horn, 2015). Cl has reciprocal connectivity with almost all cortical and subcortical brain areas as well as massive input from all major neuromodulator circuits (Smythies, Edelstein, & Ramachandran, 2012). Based on the anatomy and animal neurophysiological findings, Cl has been hypothesized as a brain network hub for multisensory integration (Mathur, 2014) and conscious percepts (Crick & Koch, 2005). It has been also hypothesized that Cl modulates cortical activities through neural synchronization, especially in high-frequency gamma band (Smythies et al., 2012; Smythies, Edelstein, & Ramachandran, 2014). As an attention theory of Cl suggested critical roles of Cl in bottom-up and top-down attention (Goll, Atlan, & Citri, 2015), recent animal studies demonstrated that Cl mediated cortical top-down control of sensory and association cortices in guiding behaviors (White & Mathur, 2018; White et al., 2018).

Although the precise functions of human Cl is largely unknown, numerous studies have suggested the significance of Cl in the pathophysiology of various neuropsychiatric disorders, including epilepsy and psychosis. Given the strong reciprocal connections of Cl with cortical/subcortical motor regions and visual cortices, as well as the amygdala, Cl appears to have a critical role in epilepsy and seizures. In the amygdala kindling model of epilepsy, in which the amygdala exhibits a progressive evolution of seizures from nonconvulsive partial seizures evoked by initial stimulations into generalized convulsive seizures, claustral lesions eliminate the convulsive component (Majak, Pikkarainen, Kemppainen, Jolkkonen, & Pitkänen, 2002; Mohapel, Hannesson, Armitage, Gillespie, & Corcoran, 2000; Wada & Kudo, 1997; Wada & Tsuchimochi, 1997). Photically induced seizures in animals were effectively modified by claustral lesions (Kudo & Wada, 1995), suggesting an important role of Cl in transforming a visual afferent to convulsive seizures in photosensitive epilepsy. In human epilepsy, it has been suggested that altered consciousness or loss of memory during seizure might be caused by disruptions in Cl. Electrical stimulation of Cl reversibly disrupted consciousness in a patient with epilepsy (Koubeissi, Bartolomei, Beltagy, & Picard, 2014). Cl appears to have a critical role in severe psychopathologies, including psychosis. Severe hallucinatory symptoms have been reported in rare cases of lesions exclusively affecting the bilateral claustrum (Ishii, Tsuji, & Tamaoka, 2011; Sperner, Sander, Lau, Krude, & Scheffner, 1996). In a post-mortem brain study, bilateral claustral volume reduction was observed in the brains of people with schizophrenia (PSZ), especially in paranoid type PSZ (Bernstein et al., 2016). A volumetric MRI study found a significant correlation between claustral gray matter atrophy and delusional severity in PSZ (Cascella, Gerner, Fieldstone, Sawa, & Schretlen, 2011). Interestingly, recent quantitative meta-analysis of fMRI studies of schizophrenia identified consistent activations of the claustrum together with hub nodes of salience network, default mode notwork, and limbic system during auditory hallucinations in PSZ (van Lutterveld, Diederen, Koops, Begemann, & Sommer, 2013; Zmigrod, Garrison, Carr, & Simons, 2016). Children with autism spectrum disorder (ASD), the neurodevelopmental disorder that is highly related to schizophrenia, were found to have substantial volume reductions in Cl in a post-mortem and MRI studies too (Davis, 2008; Wegiel et al., 2015, 2014). These findings suggest that dysfunctional claustrum might underlie epileptic seizures, altered sensory experiences, hallucinations, and severe developmental problems in cognitive and social functions.

Despite accumulating evidence that the Cl may play a role in epilepsy, schizophrenia and other neuropsychiatric disorders (Meletti et al., 2015; Silva, Jacob, Melo, Alves, & Costa, 2017), there are several reasons why the Cl has not been systematically investigated in humans. First, exclusive lesions in bilateral claustrum are extremely rare, thus no neuropsychological investigation has systematically documented putative function of Cl in human behavior. Second, its thin (1-5 mm) sheet-like anatomy located in deep brain has hindered identification and delineation of Cl in vivo in humans using basic neuroimaging tools (e.g., no Cl label in most widely used neuroimaging brain atlases, no reliable method to delineate Cl, etc.). Third, conventional methods of the two major noninvasive *in vivo* brain activity measures, including functional neuroimaging (i.e., PET and fMRI) and high-density neurophysiological recordings (i.e., MEG and EEG) have limited spatial resolutions that cannot isolate Cl neural activities. Therefore, functional neuroimaging approaches have not been used to determine the role of Cl in human cognition and behavior as well as in neuropsychiatric disorders.

With the recent advancement of neuroimaging techniques including high spatial resolution MRIs, it became possible to investigate Cl using neuroimaging approaches. However, most widely used probabilistic brain atlases (e.g., Harvard-Oxford cortical and subcortical brain atlas; http://www.cma.mgh.harvard.edu/fsl_atlas.html) as well as a recent brain atlas based on connectional architecture (e.g., Brainnetome atlas; http://atlas.brainnetome.org/index.html) do not include the claustrum. Therefore, the claustrum has been largely ignored and its role in key cognitive functions have been misassigned to nearby structures (e.g., insula, putamen, amygdala, etc.) in neuroimaging studies. Furthermore, it is difficult to conduct region-of-interest (ROI) analysis of the claustrum since no neuroimaging data analysis software provides automated parcellation of the claustrum. The old version of Freesurfer provided an automatic parcellation of the claustrum, but it was removed in the later version due to its inaccuracy. The only available option for the claustrum ROI-based approach is to manually segment the claustrum using structural MR images. Earlier study developed a manual tracing protocol for a volumetric study of the human claustrum (Davis, 2008), but the protocol did not provide enough details of the unique structure of the claustrum and no clear boundary to delineate the sub-regions of the claustrum. Therefore, as the methodological foundation for future neuroimaging studies of Cl, it is imperative to develop a comprehensive and accurate method to segment Cl.

Due to its thin morphological feature lying in the center of each hemisphere, Cl is a challenging brain structure to delineate using MRIs. Also, the morphology is determined primarily by the surrounding white matter fibers whose structural plasticity is high (Bengtsson et al., 2005; Blumenfeld-Katzir, Pasternak, Dagan, & Assaf, 2011). Therefore, there are large individual differences in the morphology of Cl. Cl in most mammals has been broadly divided into dorsal and ventral parts, which are also called "insular claustrum" and "the endopiriform nucleus" in non-primate mammals (Kowiański, Dziewiątkowski, Kowiańska, & Moryś, 1998; Reser et al., 2014). The dorsal and ventral Cl are distinguished by cell number and density, connections, and calcium binding proteins (Kowiański et al., 1998) as well as their distinct connection profiles (i.e., dorsal Cl (Cld) prominently connected with cerebral cortices and ventral claustrum (Clv) with strong connections with subcortical areas. In neurosurgical and tractographic anatomical studies, Fernandez-Miranda and colleagues further classified Clv into superior and inferior parts (Fernández-Miranda, Rhoton, Kakizawa, Choi, & Alvarez-Linera, 2008; Pathak & Fernandez-Miranda, 2014).

The present work provides a comprehensive set of protocol for manually tracing the human claustrum. We used high resolution (0.7 mm$^3$ isotropic voxel size) MRI that is available from *the Washington University-Minnesota Consortium Human Connectome Project (WU-Minn HCP)* (David C. Van Essen et al., 2013). The protocol was developed based on a cellular-level human brain atlas (Ding et al., 2017) that incorporated neuroimaging (T1- and diffusion-weighted MRI), high-resolution histology, large-format cellular resolution (1 μm/pixel) Nissl and immunohistochemistry anatomical plates of a complete adult brain. The protocol includes detailed descriptions for delineating three sub-regions of Cl including dorsal, ventral, and temporal claustrum (Clt) that were identified based on cyto- and chemoarchitecture of the brain atlas. The ventral and temporal claustrum in our protocol correspond to the superior and inferior parts of Clv that Fernandez-Miranda and colleagues identified in their studies.

## Methods

*Subjects*

To understand inter-individual variability of brain circuits and its genetic bases and relation to behavior, the WU-Minn HCP subjects were drawn from a population of young adults (age range 22 – 35 years) twins and their non-twin siblings. The HCP subjects represent healthy adults that had no prior history of significant neurological and psychiatric illnesses that were determined based on semistructured interviews and comprehensive neurological and psychological assessments. Any sibships with individuals with severe neurodevelopmental, neuropsychiatric, and neurological disorders were excluded. The details of the HCP subjects are described elsewhere (D C Van Essen et al., 2012; David C Van Essen et al., 2013). For the claustrum segmentation protocol development, we randomly selected brain scans of 10 human subjects (2/2 males/females, age range 22–35) from the WU-Minn HCP.

*MRI Scan Acquisition and Processing*

All subjects were scanned on a customized Siemens 3T Connectome Skyra scanner. Structural images were acquired using the 3D MPRAGE T1-weighted sequence with 0.7 mm isotropic resolution (FOV=224 mm, matrix=320, 256 sagittal slices in a single slab, TR=2400 ms, TE=2.14 ms, TI=1000 ms, flip angle=8°). The T1 MRI was preprocessed following the HCP protocol that is described elsewhere (Glasser et al., 2013).

*Segmentation protocol*

We used 3DSlicer/v.4.5.1 software (http://www.slicer.org) and the Web-based cellular-level brain atlas of the adult human (http://atlas.brain-map.org/; Ding et al., 2017) with a high user-interactivity for zooming, highlighting a specific annotation region, etc. The brain atlas provides comprehensive details of structural annotation for 862 brain structures in 106 coronal plates, which is oblique to the AC-PC plane to which the WU-Minn HCP T1-MRI data is aligned. In particular, the majority of the anterior portions of the Clt that are most challenging to trace are visible in the Slab 4 slices, which have the oblique angle of 10 degrees (see Supplementary Figure 1). To align the HCP T1-MRI data to the coronal view plane of the brain atlas, we rotated the T1-MRIs by 10 degrees along the AC-PC axis using the Transforms module of the Slicer. For accurate translation of anatomical landmark descriptions of the brain atlas in the claustrum tracing, we created the new rotated MRI volumes with the same isotropic voxel sizes by re-gridding with the Slicer Crop Volume module using linear interpolation. To ensure equivalent contrast range across the claustrum segmentations, T1 images were adjusted by setting the display window width/level (brightness/contrast) to 500 and 700 that highlight the claustrum nuclei.

Subcortical structures that are often segmented by a manual tracing approach (e.g., amygdala, hippocampus, caudate, etc.) primarily uses the coronal planes. However, Cl whose complex morphology is determined by nearby white matter and gray matter structures requires a more sophisticated tracing strategy that uses different planes for the different sub-regions. For example, it is easiest to trace Cld in the axial planes, while it is easiest to trace Clv in the coronal planes. If one continues to use only a single view for tracing the entire claustrum, it is easy to trace erroneously adjacent cortical and subcortical regions having transitional regions that appear to be connected to the claustrum in the view. Therefore, it is highly recommended to trace the claustrum following the order in the specified planes as described below. Also, it is critical to correct any erroneous tracing with the final visual check using the three views.

*Tracing the dorsal claustrum (Cld)*

Tracing begins in the axial plane starting from around the bottom of the central anterior commissure (AC) where the thin morphology of Cld lying between the putamen and the insular cortex is clearly visible (see Figure 1). Cld is elongated across the anterior-posterior axis and medially separated from the putamen by the external capsule and laterally separated from the insular cortex by the extreme capsule. The rostral, caudal, dorsal ends of Cld are approximately the same of those of the putamen. Cld narrows in the superior directions, so tracing dorsally until the narrow Cld nuclei around the dorsal boundary is not distinguishable. Tracing the dorsal ends of Cld should be completed in the subsequent tracing of Clv in the coronal planes that show the dorsal boundaries more clearly.

*Tracing the ventral claustrum (Clv)*

Clv is clearly defined in the coronal view where it is well separated by the external and extreme capsules (see Figure 2). Starting from the coronal plane intersecting around the middle of Cl in an axial plane, trace anteriorly until the claustrum nuclei are indistinguishable. Then go back to the starting point and complete to trace the posterior Clv. In this step, trace only Clv that is superior to the temporal lobe. In the coronal view, the anterior Clv extends inferior to the putamen following the exterior curvature of the putamen. The anterior Clv is clearly separated dorsomedially from the putamen by the external capsule, while it is separated ventrolaterally from the base of the frontal lobe including orbitofrontal cortex and frontal agranular insular cortex by the extreme capsule. But, the inferior border of the anterior Clv partly adjoins to the orbitofrontal cortex (OFC) and the frontal agranular insular cortex (FI). Thus, it needs to trace the anterior Clv carefully not to include parts of the orbitofrontal and the insular cortices adjoining to the inferior border of Clv. They are distinguishable in the coronal planes. The rostral and caudal borders of Clv are better identified in the axial view, so it needs to finalize them in the axial view in the final visual check.

*Tracing the temporal claustrum (Clt)*

The inferior parts of Clv extend toward the temporal lobe, forming Clt. In the anterior-to-posterior axis, Clt ranges from the rostral tip of the endopiriform nucleus (EN) to the rostral tip of the hippocampus (see Figure 2.iv and 2.vii that shows the rostral and the caudal ends of tCl). The anterior part of Clt is the most difficult region of Cl to trace because it is surrounded by the gray matters of frontal and temporal lobes and adjoins to the amygdaloidal complex, especially to EN. The relationship between the amygdaloidal complex and Clt is so close that the anterior Clt is often called as the periamygdalar claustrum (Zelano & Sobel, 2005). The ventrolateral border of the anterior Clt adjoins to the temporal angular insular cortex (TI) and the entorhinal cortex (EC; in the rostral part) and white matter bundles including uncinated fasciculus (UF) and inferior occipitofrontal fasciculus (IFOF; in the caudal part). The dorsomedial border of the anterior Clt adjoins to EN and the piriform cortex (Pir). Due to multiple transitional areas between the anterior Clt and the adjoining regions, it is hard to delineate the clear borders of the anterior Clt. Despite the challenging anatomical features of the anterior Clt, the HCP high resolution MRIs with 0.7mm$^3$ isotropic voxel sizes allows one to delineate the Clt nuclei that looks darker than the surrounding gray matters. As shown in the Nissl-stained brain slices in Figure 2, Clt has dense gray matter nuclei standing out in the background of the adjacent cortical and subcortical structures in the medial temporal lobe region.

To trace the anterior Clt, it needs to use the coronal and the axial planes simultaneously in a way that the primary tracing in the coronal planes is followed by adjustments and confirmations in the corresponding axial planes. The coronal view is useful to identify the overall morphology and the axial view is useful to identify the border more accurately (see Figure 3.i and 3.ii). In this procedure, it needs to use a side-by-side display of the two planes and a navigation tool of a crosshair showing the intersections across the two panels. It needs to use extra care during the procedure to include only the anterior Clt that is darker than the surrounding gray matters. As the first step, it needs to identify EN as the useful landmark that guides to delineate the medial border of the anterior Clt in the coronal view. The rostral tip of EN is located around the junction between the frontal and the temporal lobes (marked by a green arrow in Figure 2 panel iv). Once the EN landmark point is localized in the coronal plane, begin to trace the anterior Clt that adjoins to the dorsolateral side of the landmark point and extends toward the ventromedial direction. Proceed to trace toward the posterior Clt that is embedded in the white matter bundles in the temporal lobe.

The posterior Clt is relatively easier to trace in the coronal planes than the anterior Clt. However, the white matter fibers of UF and IFOF traversing the posterior Clt, so it is fragmented into many small pieces. Therefore, it is required to supplement the coronal view tracing using the sagittal view that clearly shows the main body of the lateral posterior Clt (see Figure 3.iii). It is noteworthy that one should not use the sagittal planes for refining the medial border of the Clt. In the sagittal view, it is very hard to identify the transition from the medial border of the anterior Clt to the lateral border of the amygdala.

It is required to conduct the final visual check and make corrections using multiple planes. Figure 3 shows the axial and the sagittal planes, in which the brain regions adjoining to Clt (EN, amygdalostriatal transition area [ASTA], and TI) are depicted in green color. It also highlights the adjacent regions that are hard to be differentiated from Clt, including posteroventral putamen (PuPV) in the coronal planes. It is also recommended to look up the online brain atlas (http://atlas.brain-map.org/) atlas to obtain additional guidance for the anatomical details of Cl that are not fully covered by the current protocol.

*Parcellation of Cl into the three sub-regions*

As described so far, Cl is separated into the three sub-regions across the superior-to-posterior axis. A geometric method can be used to parcellate them accurately. The first step is to identify three key points on the claustrum tracing in the coronal planes (Figure 4). Point A is the superior-lateral point of the temporal lobe. Point B is the most inferior point in the putamen. Point C is identified as the most lateral point of the outer curvature of the putamen. Point A and C should be identified across the entire coronal planes where Clv is identified. Point B should be identified only for the coronal planes covering the anterior-to-posterior ranges of Clt (i.e., from the coronal plane showing the rostral tip of EN to the coronal plane showing the rostral tip of the hippocampus). The second step is to draw two straight lines, including Line 1 connecting Point A and B and Line 2 connecting Point A and C. All Cl above Line 1 is Cld, all Cl below Line 2 is Clt, and the remaining is Clv. The parcellation rule has one exception; if any small branches of Cl that hook ventrolaterally cross Line 1 or 2, the branches should be parcellated according to the membership of the main Cl body region from which the branches originate.

## Representative Results and Future Directions

For illustration purposes, a 3D model of the manual segmentation for the whole Cl of one subject is shown in Figure 5. As depicted in a background of a sagittal plane, Cl is large in its anterior-posterior and dorsal-ventral extent although it looks small in coronal and axial planes. The volumes of the whole, dorsal, ventral, and temporal Cl were 3.43, 0.51, 1.48, and 1.44 $cm^3$, respectively. Given the average volume of the amygdala estimated in an MRI volumetric study that manually segmented the 30 adult individuals (3.26 $cm^3$; Brabec et al., 2010) Cl is a fairly large subcortical nucleus. It is important to assess the reliability of the manual segmentation protocol that can be used for future anatomical studies of various neuropsychiatric disorders, including epilepsy and schizophrenia. The comprehensive and reliable manual segmentation method will contribute to a useful tool for identifying Cl in human MRIs, such as a probabilistic brain atlas of Cl and an automatic Cl segmentation software.

**Box 1. Abbreviations**

> AC, anterior commissure; ASTA, amygdalostriatal transition area; BF, basal forebrain; Ca, caudate nucleus; Cl, the claustrum; Cld, the dorsal claustrum; Clv, the ventral claustrum; Clt, the temporal claustrum, CIdg, caudal dysgranular insular cortex; EC, entorhinal cortex; En, endopiriform nucleus; FI, frontal agranular insular cortex; GP, globus pallidus; Iag, agranular insular cortex; Ins, insular cortex; LOA, lateral olfactory area; PI, parainsular cortex; Pir, piriform cortex; Pu, putamen; PuPV, posteroventral putamen; Ridg, rostral dysgranular insular cortex; OFC, orbitofrontal cortex; TI, temporal angular insular cortex.

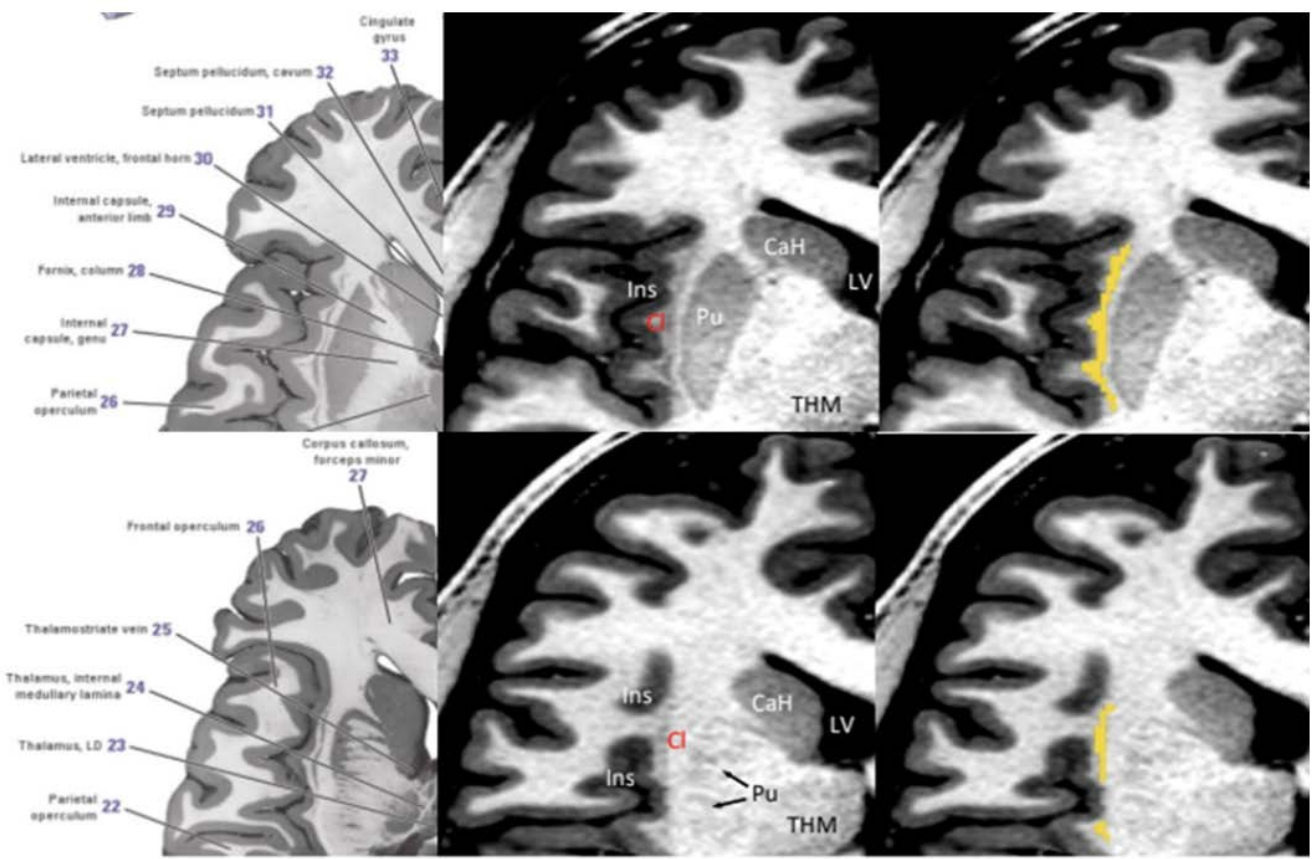

Figure 1. Cld in the axial planes. Cld lies between the putamen (Pu) and the insular cortex (Ins). Cld tracing is

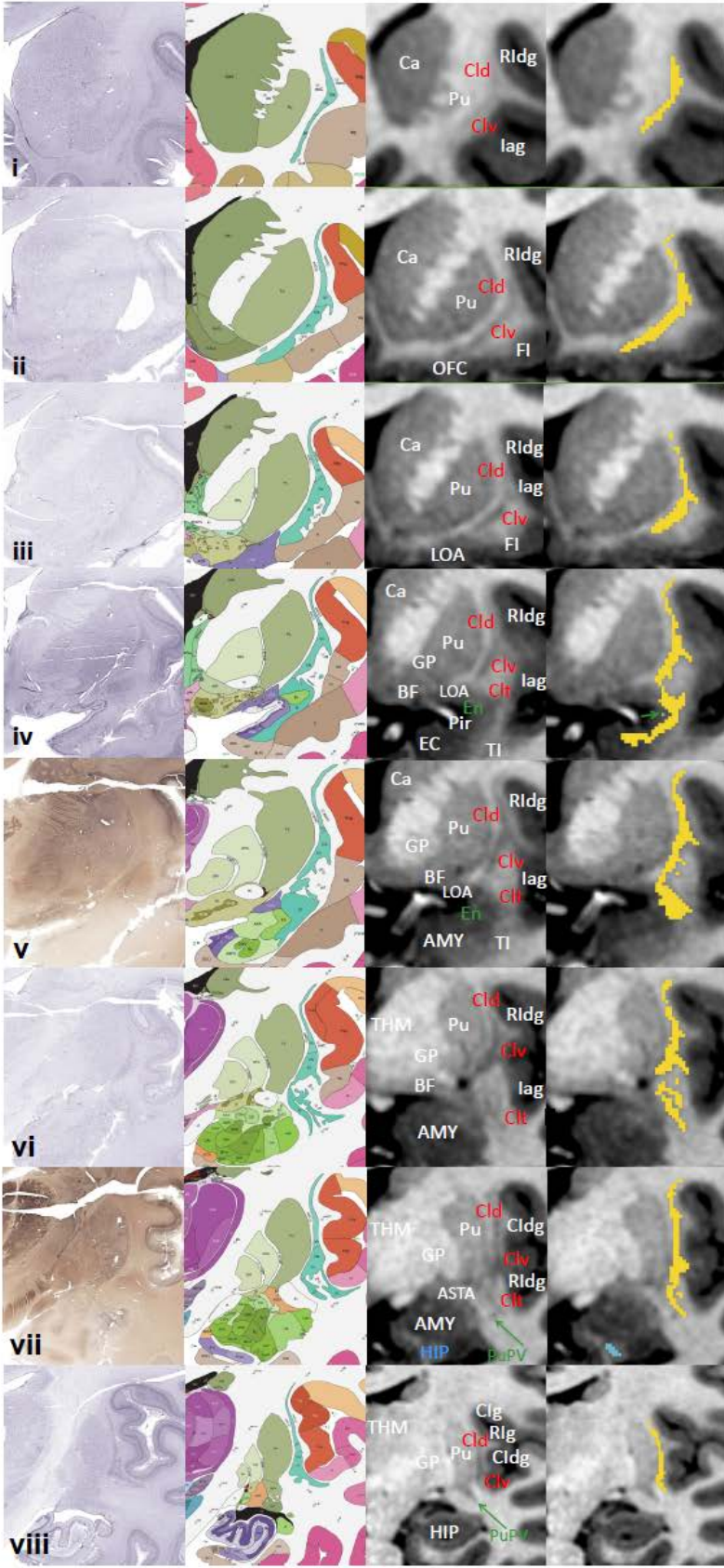

Figure 2.

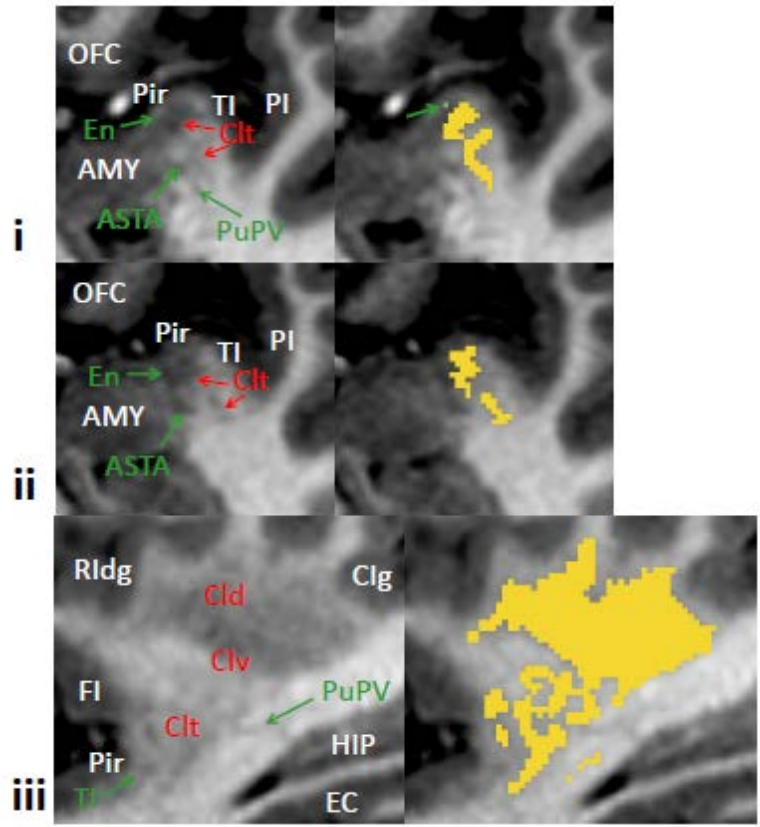

Figure 3.

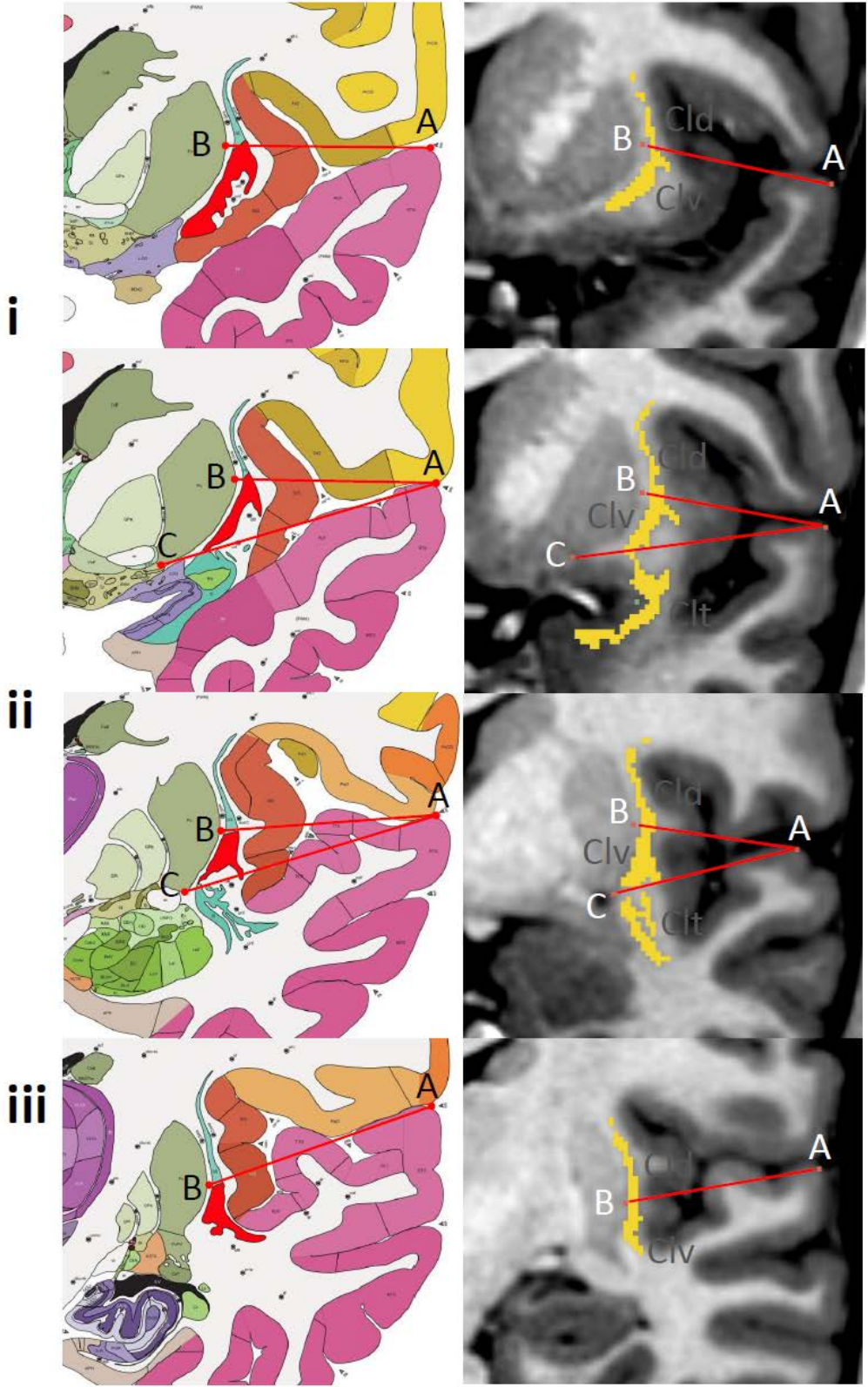

Figure 4.

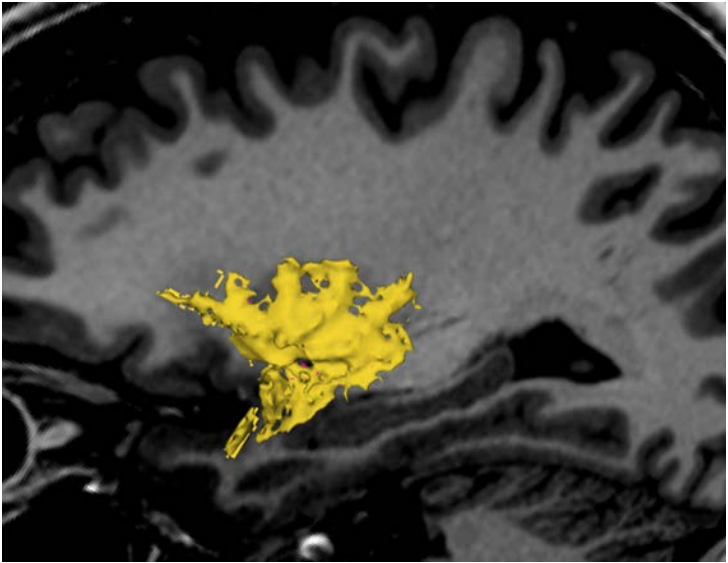
Figure 5.

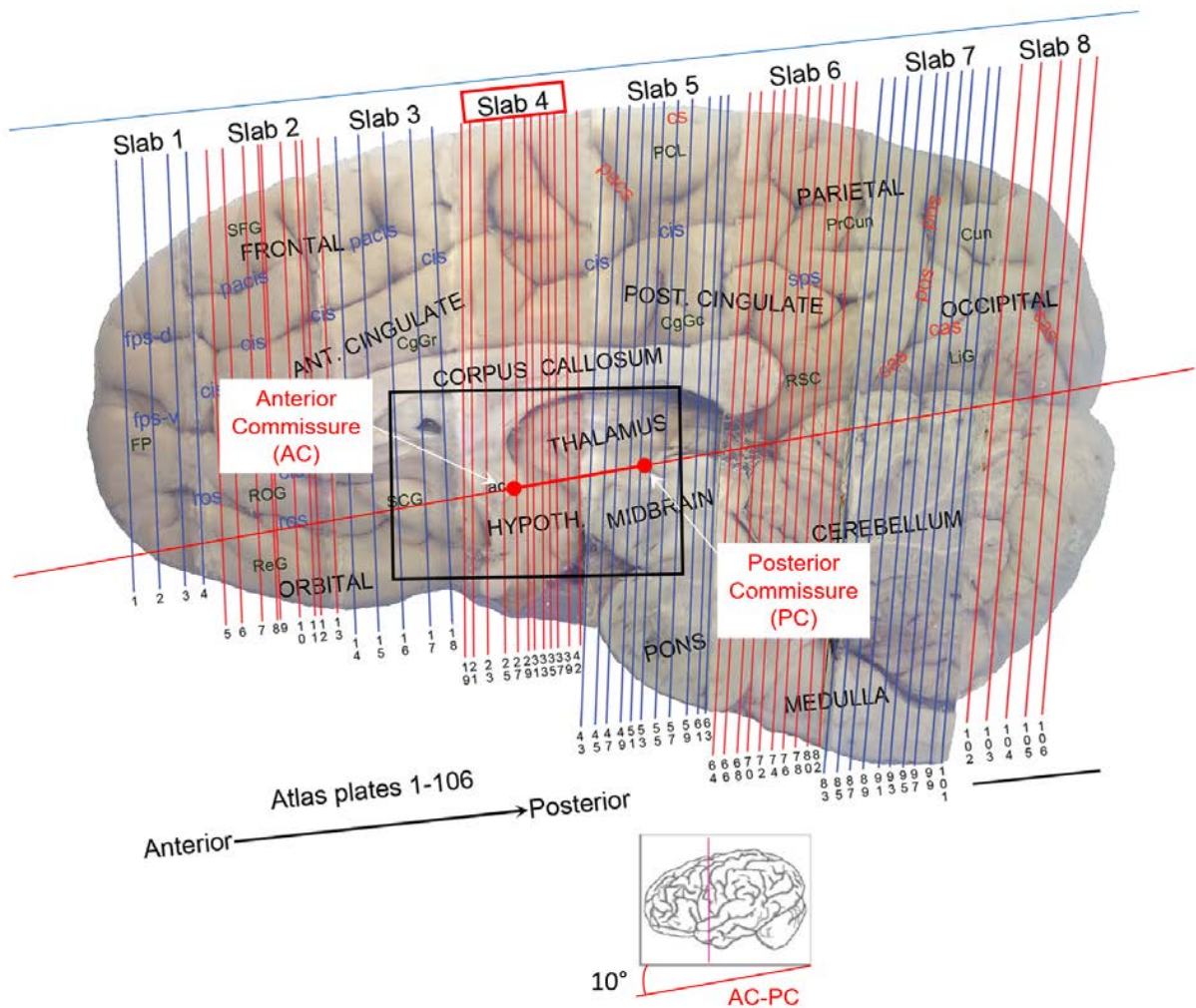

**Supplementary Figure 1. The Allen human brain atlas with the coronal plates oblique to the AC-PC plane.** The central rectangle highlights the area where the claustrum is visible in the brain atlas (plates 16-57). In particular, the Slab 4 plates where the anterior portions of the temporal claustrum are included have the oblique angle of 10 degree to the AC-PC axis, which requires a rotation of the AC-PC aligned structural scans for accurate tracing of the temporal claustrum based on the brain atlas (adapted from Figure 16 of Ding et al., 2017).